\documentclass[12pt,amsmath,amssymb,]{revtex4}
\usepackage{amsfonts}
\usepackage{amsmath}
\usepackage{amssymb}
\usepackage{array}
\usepackage{bm}
\usepackage{braket}
\usepackage{breqn}
\usepackage{color}
\usepackage{dcolumn}
\usepackage{epsfig}
\usepackage{epsf}
\usepackage{epstopdf}
\usepackage{float}
\usepackage{graphicx}
\usepackage{graphics}
\usepackage{harpoon}
\usepackage{indentfirst}
\usepackage{mathrsfs}
\usepackage{multirow}
\usepackage{pifont}
\usepackage{xcolor}

\newcommand{\RM}[1]{\textrm{\uppercase\expandafter{\romannumeral#1}}}

\begin{document}

\title{ Recoil contributions in neutron $\beta^-$ decay and their corrections to the correlation coefficients}

\author{
Hui-Yun Cao$^{1}$\protect\footnotemark[1]\protect\footnotetext[1]{E-mail: caohy@hbnu.edu.cn},
Hai-Qing Zhou$^{2}$\protect\footnotemark[2]\protect\footnotetext[2]{E-mail: zhouhq@seu.edu.cn} \\
$^1$ School of Physics and Electronic Science, Hubei Normal University, HuangShi 435002, China\\
$^2$ School of Physics, Southeast University, NanJing 211189, China}

\date{\today}

\begin{abstract}
In this work, the recoil corrections to the correlation coefficients in the neutron $\beta^-$ decay with polarized neutron and electron are evaluated to order $O(m_n^{-2})$ in the Born approximation. Different from the usual calculations applied in the literatures where the phase space factor and the amplitude are expanded independently, we directly expand the full differential scattering cross section and then integrate the solid angles of the neutrino. Furthermore, the most general coupling forms in $Wpn$ are kept in the calculation and the analytic expressions for the angular correlation coefficients are given. The numerical comparisons between the results by the different treatments of the phase space factor and the different choices of the coupling parameters are also presented. The results show that the careful treatment of the expansion on $m_n^{-1}$ and the inclusion of the general coupling forms are necessary when aiming to reach the precision $10^{-5}$.

\end{abstract}

\maketitle

\section{Introduction}


The $\beta^-$ decay of a free neutron $n\rightarrow pe\bar{\nu}_e$ \cite{Abele-2008} provides a clean process to determine the elemental parameters in the standard model (SM).  In the SM, a free neutron is unstable and its $\beta^-$ decay is mainly governed by the weak interaction. In this process, the Cabibbo-Kobayashi-Maskawa(CKM) matrix element $V_{ud}$ and the weak coupling of $Wpn$  play the roles. Due to the clean background, it is a powerful laboratory to check the universality of the quark mixing CKM matrix \cite{Abele-2002}, the conserved-vector-current(CVC) hypothesis \cite{CVC-hypothesis}, as well as the absence of second-class currents(SCC) \cite{SCC}, {\it etc}. In addition, the axial-vector coupling constant $g_A$ provides a necessary input for nuclear physics, particle physics, and cosmology, as well as astrophysics \cite{gA-role}.

For the unpolarized $\beta^-$ decay of neutron, the experimental observed variables are the decay lifetime $\tau_n$ and the spectrum shape. Furthermore, when considering the $\beta^-$ decay with a polarized electron and a polarized neutron, the angular correlation coefficients $A(E_e), G(E_e), N(E_e), Q(E_e)$ and $R(E_e)$ can also be measured \cite{correlation-coefficients-AGNQR}. Recently, many high precise measurements have been carried out in the neutron $\beta^-$ decay, including the non-polarization case \cite{Experiments-unpolarized}, the single neutron polarization \cite{Experiments-neutron-polarized}, and both the electron and the neutron polarizations \cite{Experiments-neutron-and-electron-polarized}. To extract the accurate value of $V_{ud}$ and the coupling constants from the free neutron $\beta^-$ decay, the theoretical estimations for the corresponding correlation coefficients should reach the sufficient accuracy. For instance, the model-independent radiative corrections (MIRC) in unpolarized case have been evaluated to the leading-order $O(\alpha_{\text{QED}})$ in \cite{MIRC-leading-order}, the next-to-leading-order $O(\alpha_{\text{QED}}^2)$ in \cite{MIRC-to-NL-order}, and the next-to-next-to-leading-order $O(\alpha_{\text{QED}}^3)$ in \cite{MIRC-to-NNL-order}, with $\alpha_{\text{QED}}$ being the fine structure constant. Furthermore, the model-dependent radiative corrections (MDRC) are estimated by many different methods, such as a renormalization group analysis method \cite{MDRC-renormalization-group-Serlin-1986} and the improved method \cite{MDRC-improved-Serlin-2006}, the effective field theory method \cite{MDRC-EFT-Ando}, and the dispersion relation method \cite{MDRC-DR-Chien}. Other contributions from the proton recoil, the finite proton radius, and the lepton-nucleon convolution \cite{other-contribution-recoil-proton-radium-convolution} are also non-negligible for the upcoming high precision extractions \cite{Futrue-high-precision-Ex-neutron}.

In the recent literatures \cite{Ivanov-2021,Ivanov-2018}, the contributions to the correlation coefficients $A(E_e), G(E_e), N(E_e), Q(E_e)$ and $R(E_e)$ due to the proton recoil effect at the tree level are estimated to the precision $10^{-5}$. These estimations are based on the Born approximation(where only the tree diagram and the leading coupling constants are considered) and the independent expansions of the amplitude $\mathcal{M}$ and the 3-body phase factor on $m_N^{-1}$, where $m_N=(m_n+m_p)/2$ with $m_n$ and $m_p$ are the mass of neutron and proton respectively. In this work, we calculate these contributions in a general form where all the possible couplings are kept and the expansion on $m_n^{-1}$ is done for the final full results. The final analytic expressions and the numerical comparison show some differences with the results in the literatures.

The paper is organized as follows: In Sec. II we give the basic formula in the Born approximation, in Sec. III we present the analytical expressions for the angular correlation coefficients to $\mathcal{O}(m_n^{-2})$, in Sec. IV the numerical comparisons between our results and those in the literatures are presented. The carefully discussion on the differences due to the different choices of the phase factors and the additional form factors are also given.

\section{Basic Formula for $n\rightarrow p e \bar{\nu}_e$}
\begin{figure}[htbp]
\centering
\includegraphics[height=5cm]{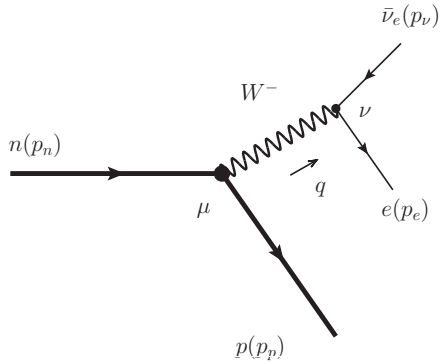}
\caption{Diagram for $n\rightarrow p e \bar{\nu}_e$ with one-W-exchange.}
\label{figure:tree-diagram}
\end{figure}

In the Born approximation where only the one-W-exchange diagram shown in Fig. \ref{figure:tree-diagram} is considered, the corresponding amplitude of the free neutron $\beta^-$ decay can be written as

\begin{eqnarray}
\mathcal{M}=-i\frac{G_F}{\sqrt{2}}V_{ud}\Big[\bar{u}(p_e,m_e)\gamma_{\mu}{\color{black}(g_{eV}-g_{eA}\gamma_5)} u(p_v,m_v) \Big] \Big[\bar{u}(p_p,m_p)\Gamma^{\mu}_{W n p}(q) u(p_n,m_n)\Big],
\label{equation:Hamiltonian}
\end{eqnarray}
where $G_F=\sqrt{2}g^2/8m_W^2$ is the Fermi weak constant ($g$ is the $SU(2)$ gauge coupling constant), $V_{ud}$ is the CKM matrix element\cite{PDG-2020}, and $g_{eV,eA}$ are the coupling constants of $e\bar{\nu}_e W^-$. $\bar{u}(p_e,m_e), u(p_v,m_v),\bar{u}(p_p,m_p)$ and $ u(p_n,m_n)$ are the spinors of the electron, antineutrino, proton and neutron with the corresponding momentum and mass, respectively, and $q=p_n-p_p$. The most general form for the vertex $\Gamma_{W n p}$ in V-A theory \cite{General-form-of-Wnp-in-VA} reads as
\begin{eqnarray}
\Gamma^{\mu}_{W n p}(q) &=&  f_1(q^2)\gamma^{\mu}{\color{black}-}i\frac{f_2(q^2)}{m_n+m_p}\sigma^{\mu\rho}q_{\rho}+\frac{f_3(q^2)}{m_n+m_p}q^{\mu} \nonumber\\
&&+\Big[ f_4(q^2)\gamma^{\mu}{\color{black}-}i\frac{f_5(q^2)}{m_n+m_p}\sigma^{\mu\rho}q_{\rho}+\frac{f_6(q^2)}{m_n+m_p}q^{\mu} \Big]\gamma_5,
\end{eqnarray}
where $\sigma^{\mu\rho}=\frac{i}{2}[\gamma^{\mu},\gamma^{\rho}]$, $f_1(q^2)$, $ f_2(q^2)$, $f_3(q^2)$, $f_4(q^2)$, $f_5(q^2)$, and $f_6(q^2)$ in $\Gamma^{\mu}_{W n p}$ account for vector, weak magnetism, scalar, axial vector, weak electricity, and induced pseudoscalar contributions, respectively.

In the Born approximation, according to the power of the low momentum transfer $q^2$, the form factors $f_i(q^2)$ can be parameterized as
\begin{eqnarray}
f_1(q^2)&=&f_1+\frac{q^2}{m_n^2}\lambda_{f_1},\nonumber\\
f_2(q^2)&=&f_2,\nonumber\\
f_3(q^2)&=&f_3,\nonumber\\
f_4(q^2)&=&f_4+\frac{q^2}{m_n^2}\lambda_{f_4},\nonumber\\
f_5(q^2)&=&f_5,\nonumber\\
f_6(q^2)&=&f_6,
\end{eqnarray}
where $f_{1-6}$ and $\lambda_{f_{1},f_{4}}$ are the constants. For simplification, we denote $\lambda_{f_1}$ and $ \lambda_{f_4}$ as $f_7$ and $f_8$. In the literatures, usually only the terms $f_1,f_2,f_4$ are kept \cite{Ivanov-2013,Ivanov-2017,Ivanov-2018,Ivanov-2021}. In our discussion, we keep all these terms for generality.



The differential scattering cross section with a polarized neutron and a polarized electron can be written as
\begin{eqnarray}
d\sigma_n &=& \frac{F(E_e,Z=1)}{2E_n}\prod_{i=e,v,p} \frac{d^3 \vec{p}_i (2\pi)^4 \delta^4(p_e+p_v+p_p-p_n) }{(2\pi)^3(2E_i)} \sum_{\text{helicity of} ~p, ~\bar{v}_e} \mathcal{M}\mathcal{M}^* ,
\end{eqnarray}
where $E_i$  are the energies of the corresponding particles in the rest frame of neutron, $F(E_e,Z=1)$ is the relativistic Fermion function \cite{Fermin-function} which describes the contribution of the electron-proton final state Coulomb interaction. After integrate the $\delta$ functions, one can get
\begin{eqnarray}
\frac{d^5\sigma_n(E_e,\vec{k}_e,\vec{\xi}_e, \vec{\xi}_n)}{dE_e d\Omega_e d\Omega_v} &= & F(E_e,Z=1) \beta\sum_{\text{helicity of} ~p, ~\bar{v}_e} \mathcal{M} ~ \mathcal{M}^*,
\end{eqnarray}
where $\vec{k}_e$ is the three-momentum of electron in the rest frame of neutron, $\vec{\xi}_n$ and $\vec{\xi}_e$ are the polarization vectors of the neutron and electron, $d\Omega_e$ and $d\Omega_v$ are the elements of the solid angles of the electron and the neutrino, respectively. The 3-body phase space factor $\beta$ reads
\begin{eqnarray}
\beta&=& \frac{1}{16m_n}\frac{1}{(2\pi)^5}\frac{\sqrt{E_e^2-m_e^2}E_v}{E_p+E_v+\vec{k}_e\cdot \vec{n}_v},
\label{equation:phase-factor-beta-1}
\end{eqnarray}
where the unit vector $\vec{n}_v$ is directed along the neutrino three-momentum $\vec{k}_v$. To deal with the phase space in a convenient way, for example the authors of Ref. \cite{MDRC-EFT-Ando} expand the following variable $Y_1$ in their Eq. (20) as
\begin{eqnarray}
Y_1&\equiv&\frac{m_p E_v^2}{E_p+E_v+\vec{k}_e\cdot \vec{n}_v} \nonumber\\
&=& 16m_n(2\pi)^5 \frac{m_pE_v}{\sqrt{E_e^2-m_e^2}}\beta\nonumber\\
&\approx&(E_0-E_e)^2\Big[1+\frac{2}{m_n+m_p}(3E_e-E_0-3\vec{k}_e\cdot\vec{n}_\nu)\Big],
\end{eqnarray}
where $E_0=(m_n^2-m_p^2+m_e^2)/2m_n$ is the endpoint energy of the electron spectrum.

Similarly, if one expand  the following variable $Y_2$  on $m_n^{-1}$ to the order $\mathcal{O}(m_n^{-2})$
\begin{eqnarray}
Y_2\equiv\frac{(E_0-E_e)^2}{E_p+E_v+\vec{k}_e\cdot \vec{n}_v} = 16m_n(2\pi)^5 \frac{(E_0-E_e)^2}{\sqrt{E_e^2-m_e^2}E_v}\beta,
\end{eqnarray}
one can get
\begin{eqnarray}
\beta\rightarrow \beta_{\RM1}&=& \frac{1}{16m_n}\frac{1}{(2\pi)^5}\Big[1+\frac{3}{m_n}(E_e-\frac{\vec{k}_e\cdot\vec{n}_v}{E_e})\Big]\frac{(E_0-E_e)^2\sqrt{E_e^2-m_e^2}}{m_nE_v},
\label{equation:phase-factor-beta-2}
\end{eqnarray}
In the literatures \cite{Ivanov-2013,Ivanov-2017,Ivanov-2018}, the phase factor is approximately taken as
\begin{eqnarray}
\beta\rightarrow\beta_{\RM2}&=& \frac{1}{16m_n}\frac{1}{(2\pi)^5}\Big[1+\frac{3}{(m_n+m_p)/2}(E_e-\frac{\vec{k}_e\cdot\vec{n}_v}{E_e})\Big]\frac{(E_0-E_e)^2\sqrt{E_e^2-m_e^2}}{m_nE_v},
\label{equation:phase-factor-beta-3}
\end{eqnarray}
which has a factor difference with  $\beta_{\RM1}$. In our numeric calculations, these three phase factors $\beta,\beta_{\RM1}$ and $\beta_{\RM2}$ are all used for comparison to show the difference.

Experimentally, the direction of the three-momentum of the $\bar{\nu}_e$ is not measured, and so we should integrate over $d\Omega_v$. After the integration of $d\Omega_v$, the form of the expression can be written as {\color{black}\cite{correlation-coefficients-AGNQR,Ivanov-2017}},
\begin{eqnarray}
\frac{d^3\sigma_n(E_e,\vec{k}_e,\vec{\xi}_e, \vec{\xi}_n)}{dE_e d\Omega_e} &= &  \frac{G_F^2|V_{ud}|^2}{8\pi^4}  (E_0-E_e)^2\sqrt{E_e^2-m_e^2}E_e F(E_e,Z=1)\nonumber\\
&&\times \Big\{ \zeta(E_e)+ \bar{A}(E_e) ~\frac{\vec{\xi}_n\cdot\vec{k}_e}{E_e} +\bar{G}(E_e)~\frac{\vec{\xi}_e\cdot\vec{k}_e}{E_e}+\bar{N}(E_e)~\vec{\xi}_n\cdot\vec{\xi}_e \nonumber\\ &&+\bar{Q}(E_e)~\frac{(\vec{\xi}_n\cdot\vec{k}_e)(\vec{\xi}_e\cdot\vec{k}_e)}{E_e(E_e+m_e)}+\bar{R}(E_e)~\frac{\vec{\xi}_n\cdot(\vec{k}_e\times\vec{\xi}_e)}{E_e}\Big\}.
\label{equation:cross-section-in-correlation-coefficients-form}
\end{eqnarray}
Comparing with the expressions in Ref. {\color{black}\cite{correlation-coefficients-AGNQR,Ivanov-2017}}, here we define $\bar{A}(E_e)=A(E_e)\zeta(E_e)$, $\bar{G}(E_e)=G(E_e)\zeta(E_e)$, $\bar{N}(E_e)=N(E_e)\zeta(E_e)$, $\bar{Q}(E_e)=Q(E_e)\zeta(E_e)$, $\bar{R}(E_e)=R(E_e)\zeta(E_e)$, respectively.

To get the expressions of $\bar{A}(E_e)$, $\bar{G}(E_e)$, $\bar{N}(E_e)$, $\bar{Q}(E_e)$ and $\bar{R}(E_e)$,  in the literatures the phase space factor and the amplitude $\cal{M}$ usually are expanded  on $m_n^{-1}$(or $m_N^{-1}$) independently such as Eqs. (\ref{equation:phase-factor-beta-2}, \ref{equation:phase-factor-beta-3}). In our calculation, we expand the full expression $\frac{d^5\sigma_n(E_e,\vec{k}_e,\vec{\xi}_e, \vec{\xi}_n)}{dE_e d\Omega_e d\Omega_v}$ on $m_n^{-1}$ directly. Some details on the expansion are described around the Eq. (\ref{Eq: replace-rules}).  After the expansion, the relevant integrations of $d\Omega_v$ can be done easily and the corresponding expressions are presented in Appendix B.

\section{Analytic Expressions}
 The general recoil corrections to the angular correlation coefficients can be written as
\begin{eqnarray}
X(E_e)&=&  \sum_{i=1}^6 \sum_{j=i}^8 \mathcal{C}_{ij}^{X} L_{ij}^{X} f_i f_j,
\end{eqnarray}
where $X$ refers to $\zeta, \bar{A}, \bar{G}, \bar{N}, \bar{Q}$, $L_{ij}^{X}$ refers to the factor from the lepton part which is $g_{eV}^2+g_{eA}^2$ or $g_{eV}g_{eA}$, ${\cal C}_{ij}^{X}$ refers to the contributions from the hadron part. In the practical calculation, the terms $\mathcal{C}^{X}_{i7}$ and $\mathcal{C}^{X}_{i8}$ with $i = 2, 3, 5, 6$ are at the order $\mathcal{O}(m_n^{-3})$ which are neglected. Furthermore, when only the OBE contribution is considered, one has $\bar{R}(E_e)=0$ and our direct calculation also shows this property.

In the practical calculation, we use the FeynCalc\cite{FeynCalc-9.3} to do the trace of Dirac matrices in the 4-dimension. To expand the result on $m_n^{-1}$ in a consistent way, we do the following three replacements in the order of
\begin{eqnarray}
E_v &\rightarrow& \frac{m_n(E_0-E_e)}{m_n - E_e + \vec{k}_e \cdot  \vec{n}_v}, \nonumber\\
m_n+m_p&\rightarrow& xm_n,  \nonumber\\
m_p&\rightarrow&\sqrt{m_n^2+m_e^2-2E_0m_n}.
\label{Eq: replace-rules}
\end{eqnarray}
Such replacements can absorb $m_p$ and avoid the mixing of the power due the existence of $m_p$. Then we expand the full expression $d^5\sigma_n$ on $m_n^{-1}$ and integrate $d \Omega_\nu$, the analytic expressions for $X(E_e)$ can be got and the final expressions are showed in the Tables \ref{table:zeta-ana}-\ref{table:Qbar-ana}. The expressions of ${\cal C}_{ij}^{X}$ at the order $\mathcal{O}(m_n^{0}), \mathcal{O}(m_n^{-1}), \mathcal{O}(m_n^{-2})$ are shown in the third column to the last column in each table, respectively. Some contributions such as $\mathcal{C}_{16}^{\zeta}$ and $\mathcal{C}_{13}^{\bar{A}}$ are zero and have been omitted in these tables. The analytical results show a general property that the contributions from $f_1^2, f_1f_4, f_4^2$  are dominant.

\begin{table}[htbp]
\centering
\begin{tabular}{p{1cm} p{2cm} p{2cm} p{4cm} p{6cm}}
  \hline\hline
   & $L_{ij}^{\zeta}$& $\mathcal{O}(1)$ & $\mathcal{O}(m_n^{-1})$ &$\mathcal{O}(m_n^{-2})$ \\
  \hline$\mathcal{C}_{11}^{\zeta}$ & $g_{eV}^2+g_{eA}^2$ & $\frac{1}{2}$ &$\frac{E_e}{m_n}$ & $\frac{(3E_0^2-8E_0E_e+32E_e^2)+(2E_0/E_e-11)m_e^2}{12m_n^2}$ \\
        $\mathcal{C}_{12}^{\zeta}$ & $g_{eV}^2+g_{eA}^2$ & 0 &0 & $\frac{(6E_0^2-16E_0E_e+16E_e^2)+(7E_0/E_e-13)m_e^2}{6m_n(m_n+m_p)}$  \\
        $\mathcal{C}_{13}^{\zeta}$ & $g_{eV}^2+g_{eA}^2$ & 0 &$\frac{m_e^2/E_e}{m_n+m_p}$ & $\frac{(-E_0/E_e+5)m_e^2}{2m_n(m_n+m_p)}$  \\
        $\mathcal{C}_{14}^{\zeta}$ & $g_{eV}g_{eA}$ & 0 &$\frac{2(E_0-2E_e)+2m_e^2/E_e}{m_n}$ & $\frac{8(4E_0E_e-7E_e^2)+8(-E_0/E_e+4)m_e^2}{3m_n^2}$ \\
        $\mathcal{C}_{15}^{\zeta}$ & $g_{eV}g_{eA}$ & 0 & 0& $-\frac{2(E_0^2-2E_0E_e)+2E_0/E_em_e^2}{m_n(m_n+m_p)}$ \\
        $\mathcal{C}_{17}^{\zeta}$ & $g_{eV}^2+g_{eA}^2$ & 0 &0 & $\frac{4(E_0E_e-E_e^2)+(2E_0/E_e+1)m_e^2}{3m_n^2}$ \\
        $\mathcal{C}_{22}^{\zeta}$ & $g_{eV}^2+g_{eA}^2$ & 0 & 0& $\frac{(3E_0^2-10E_0E_e+10E_e^2)+(4E_0/E_e-7)m_e^2}{3(m_n+m_p)^2}$ \\
        $\mathcal{C}_{24}^{\zeta}$ & $g_{eV}g_{eA}$ & 0 & $\frac{2(E_0-2E_e)+2m_e^2/E_e}{m_n}$& $\frac{8(4E_0E_e-7E_e^2)+8(-E_0/E_e+4)m_e^2}{3m_n^2}$ \\
        $\mathcal{C}_{25}^{\zeta}$ & $g_{eV}g_{eA}$ & 0 &0 & $-\frac{2(E_0^2-2E_0E_e)+2E_0/E_em_e^2}{m_n(m_n+m_p)}$  \\
        $\mathcal{C}_{33}^{\zeta}$ & $g_{eV}^2+g_{eA}^2$ & 0 & 0 & $\frac{m_e^2}{2(m_n+m_p)^2}$\\
        $\mathcal{C}_{44}^{\zeta}$ & $g_{eV}^2+g_{eA}^2$ & $\frac{3}{2}$ &$-\frac{(E_0-5E_e)+m_e^2/E_e}{m_n}$  & $-\frac{(3E_0^2+56E_0E_e-176E_e^2)+(-14E_0/E_e+77E_e)m_e^2}{12m_n^2}$ \\
        $\mathcal{C}_{45}^{\zeta}$ & $g_{eV}^2+g_{eA}^2$ &0  & $-\frac{2E_0+m_e^2/E_e}{m_n+m_p}$  & $\frac{2(E_0^2-8E_0E_e)+(5E_0/E_e-3)m_e^2}{2m_n(m_n+m_p)}$ \\
        $\mathcal{C}_{46}^{\zeta}$ & $g_{eV}^2+g_{eA}^2$ & 0 & 0 & $\frac{(-E_0/E_e+1)m_e^2}{2m_n(m_n+m_p)}$\\
        $\mathcal{C}_{48}^{\zeta}$ & $g_{eV}^2+g_{eA}^2$ & 0 & 0 & $\frac{20(E_0E_e-E_e^2)+(-2E_0/E_e+11)m_e^2}{3m_n^2}$\\
        $\mathcal{C}_{55}^{\zeta}$ & $g_{eV}^2+g_{eA}^2$ & 0 & 0 & $\frac{(6E_0^2-4E_0E_e+4E_e^2)+(4E_0/E_e-1)m_e^2}{6(m_n+m_p)^2}$\\
  \hline\hline
  \end{tabular}
\caption{Analytical results for $\mathcal{C}_{ij}^{\zeta}$, where $L_{ij}^{\zeta}$ refers to the factor from the lepton part. Some contributions such as $\mathcal{C}_{16}^{\zeta}$ and $\mathcal{C}_{23}^{\zeta}$ are zero and have been omitted in the table.}
\label{table:zeta-ana}
\end{table}
\begin{table}[htbp]
\centering
\begin{tabular}{p{1cm} p{2cm}p{2cm} p{4cm} p{6cm}}
  \hline\hline
    & $L_{ij}^{\bar{A}}$ & $\mathcal{O}(1)$ & $\mathcal{O}(m_n^{-1})$ &$\mathcal{O}(m_n^{-2})$ \\
  \hline$\mathcal{C}_{11}^{\bar{A}}$ & $g_{eV}g_{eA}$ & 0 & $\frac{-2(E_0-E_e)}{3m_n}$ &$\frac{-(3E_0^2+8E_0E_e-8E_e^2)+3m_e^2}{6m_n^2}$\\
  $\mathcal{C}_{12}^{\bar{A}}$ &$g_{eV}g_{eA}$ & 0 & $-\frac{4(E_0-E_e)}{3(m_n+m_p)}$  & $\frac{-2(E_0^2+2E_0E_e)+6m_e^2}{3m_n(m_n+m_p)}$ \\
  $\mathcal{C}_{14}^{\bar{A}}$& $g_{eV}^2+g_{eA}^2$ &-1 & $-\frac{E_0+2E_e}{3m_n}$ & $-\frac{8(E_0E_e-E_e^2)}{3m_n^2}$ \\
  $\mathcal{C}_{15}^{\bar{A}}$&$g_{eV}^2+g_{eA}^2$ & 0  & $\frac{2(E_0-E_e)}{3(m_n+m_p)}$ & $\frac{2(E_0^2+4E_0E_e-8E_e^2)-3m_e^2}{6m_n(m_n+m_p)}$   \\
  $\mathcal{C}_{16}^{\bar{A}}$& $g_{eV}^2+g_{eA}^2$&  0 &  0  & $\frac{m_e^2}{2m_n(m_n+m_p)}$  \\
  $\mathcal{C}_{18}^{\bar{A}}$& $g_{eV}^2+g_{eA}^2$ & 0 &  0  & $\frac{-(4E_0E_e-4E_e^2+3m_e^2)}{3m_n^2}$  \\
  $\mathcal{C}_{22}^{\bar{A}}$& $g_{eV}g_{eA}$ &  0 &   0 & $\frac{-(E_0^2-6E_0E_e+8E_e^2)+3m_e^2}{3m_n(m_n+m_p)}$ \\
  $\mathcal{C}_{24}^{\bar{A}}$& $g_{eV}^2+g_{eA}^2$&  0 &  $\frac{-2(2E_0-5E_e)}{3(m_n+m_p)}$ & $\frac{2(E_0^2-28E_0E_e+48E_e^2)-15m_e^2}{6m_n(m_n+m_p)}$  \\
  $\mathcal{C}_{25}^{\bar{A}}$&$g_{eV}^2+g_{eA}^2$ & 0  &  0  & $\frac{2(E_0^2-2E_0E_e-2E_e^2)}{3(m_n+m_p)^2}$ \\
  $\mathcal{C}_{34}^{\bar{A}}$&$g_{eV}^2+g_{eA}^2$ & 0  &  0  & $\frac{m_e^2}{2m_n(m_n+m_p)}$ \\
  $\mathcal{C}_{35}^{\bar{A}}$&$g_{eV}^2+g_{eA}^2$ & 0  &  0  & $-\frac{m_e^2}{(m_n+m_p)^2}$ \\
  $\mathcal{C}_{44}^{\bar{A}}$&$g_{eV}g_{eA}$      & -2 &  $\frac{4E_0-22E_e}{3m_n}$  & $\frac{(3E_0^2+40E_0E_e-136E_e^2)+21m_e^2}{6m_n^2}$ \\
  $\mathcal{C}_{45}^{\bar{A}}$& $g_{eV}g_{eA}$& 0  & $\frac{4(2E_0+E_e)}{3(m_n+m_p)}$   & $\frac{-2(E_0^2-14E_0E_e-8E_e^2)-3m_e^2}{3m_n(m_n+m_p)}$  \\
  $\mathcal{C}_{47}^{\bar{A}}$&$g_{eV}^2+g_{eA}^2$ & 0  & 0   & $-\frac{(4E_0E_e-4E_e^2+3m_e^2)}{3m_n^2}$ \\
  $\mathcal{C}_{48}^{\bar{A}}$&$g_{eV}g_{eA}$ & 0  & 0   &  $-\frac{4(8E_0E_e-8E_e^2+3m_e^2)}{3m_n^2}$\\
  $\mathcal{C}_{55}^{\bar{A}}$& $g_{eV}g_{eA}$ & 0  & 0   &  $-\frac{2E_0(E_0+2E_e)}{3(m_n+m_p)^2}$\\
  \hline\hline
\end{tabular}
\caption{Analytical results for $\mathcal{C}_{ij}^{\bar{A}}$, where $L_{ij}^{\bar{A}}$ refers to the factor from the lepton part. Some contributions such as $\mathcal{C}_{13}^{\bar{A}}$ and $\mathcal{C}_{26}^{\bar{A}}$ are zero and have been omitted in the table.}
\label{table:Abar-ana}
\end{table}
\begin{table}[htbp]
\centering
\begin{tabular}{p{1cm}p{2cm} p{2cm} p{4cm} p{6cm}}
  \hline\hline
          & $L_{ij}^{\bar{G}}$ & $\mathcal{O}(1)$ & $\mathcal{O}(m_n^{-1})$ &$\mathcal{O}(m_n^{-2})$ \\
\hline $\mathcal{C}_{11}^{\bar{G}}$ & $g_{eV}g_{eA}$ & $-1$ & $-\frac{2E_e}{m_n}$ &$\frac{-(3E_0^2-8E_0E_e+32E_e^2)+9m_e^2}{6m_n^2}$ \\
       $\mathcal{C}_{12}^{\bar{G}}$ & $g_{eV}g_{eA}$&  0 & 0 & $\frac{-2(3E_0^2-8E_0E_e+8E_e^2)+3m_e^2}{3m_n(m_n+m_p)}$\\
       $\mathcal{C}_{13}^{\bar{G}}$ &$g_{eV}g_{eA}$ &  0 & 0 & $\frac{m_e^2}{m_n(m_n+m_p)}$\\
       $\mathcal{C}_{14}^{\bar{G}}$ &$g_{eV}^2+g_{eA}^2$&  0 & $-\frac{E_0-2E_e}{m_n}$ & $-\frac{16E_0E_e-28E_e^2+3m_e^2}{3m_n^2}$ \\
       $\mathcal{C}_{15}^{\bar{G}}$ &$g_{eV}^2+g_{eA}^2$&  0 & 0 & $\frac{E_0^2-2E_0E_e}{m_n(m_n+m_p)}$\\
       $\mathcal{C}_{17}^{\bar{G}}$ &$g_{eV}g_{eA}$&  0 & 0 & $-\frac{2(4E_0E_e-4E_e^2+3m_e^2)}{3m_n^2}$\\
       $\mathcal{C}_{22}^{\bar{G}}$ &$g_{eV}g_{eA}$&  0 & 0 & $\frac{-2(3E_0^2-10E_0E_e+10E_e^2)+6m_e^2}{3(m_n+m_p)^2}$\\
       $\mathcal{C}_{24}^{\bar{G}}$ &$g_{eV}^2+g_{eA}^2$& 0  & $-\frac{E_0-2E_e}{m_n}$ & $-\frac{16E_0E_e-28E_e^2+3m_e^2}{3m_n^2}$ \\
       $\mathcal{C}_{25}^{\bar{G}}$ &$g_{eV}^2+g_{eA}^2$& 0  & 0 & $\frac{E_0^2-2E_0E_e}{m_n(m_n+m_p)}$ \\
       $\mathcal{C}_{33}^{\bar{G}}$ &$g_{eV}g_{eA}$& 0  & 0 & $\frac{m_e^2}{(m_n+m_p)^2}$\\
       $\mathcal{C}_{44}^{\bar{G}}$ &$g_{eV}g_{eA}$& $-3$  & $\frac{2(E_0-5E_e)}{m_n}$  & $\frac{(3E_0^2+56E_0E_e-176E_e^2)+27m_e^2}{6m_n^2}$ \\
       $\mathcal{C}_{45}^{\bar{G}}$ &$g_{eV}g_{eA}$& 0  & $\frac{4E_0}{m_n+m_p}$ & $-\frac{2E_0^2-16E_0E_e+3m_e^2}{m_n(m_n+m_p)}$ \\
       $\mathcal{C}_{46}^{\bar{G}}$ &$g_{eV}g_{eA}$& 0  & 0 & $\frac{m_e^2}{m_n(m_n+m_p)}$ \\
       $\mathcal{C}_{48}^{\bar{G}}$ &$g_{eV}g_{eA}$& 0  & 0 & $-\frac{2(20E_0E_e-20E_e^2+9m_e^2)}{3m_n^2}$\\
       $\mathcal{C}_{55}^{\bar{G}}$ &$g_{eV}g_{eA}$& 0  & 0 & $\frac{-2(3E_0^2-2E_0E_e+2E_e^2)+3m_e^2}{3(m_n+m_p)^2}$\\
     \hline\hline
\end{tabular}
\caption{Analytical results for $\mathcal{C}_{ij}^{\bar{G}}$, where $L_{ij}^{\bar{G}}$ refers to the factor from the lepton part. Some contributions such as $\mathcal{C}_{16}^{\bar{G}}$ and $\mathcal{C}_{23}^{\bar{G}}$ are zero and have been omitted in the table.}
\label{table:Gbar-ana}
\end{table}
\begin{table}[htbp]
\centering
\begin{tabular}{p{1cm}p{2cm} p{2cm} p{4cm} p{6cm}}
  \hline\hline
         & $L_{ij}^{\bar{N}}$ & $\mathcal{O}(1)$  & $\mathcal{O}(m_n^{-1})$ &$\mathcal{O}(m_n^{-2})$ \\
\hline  $\mathcal{C}_{11}^{\bar{N}}$ & $\frac{(g_{eV}^2+g_{eA}^2)m_e}{E_e}$  & 0&$\frac{E_0-E_e}{3m_n}$ & $\frac{3E_0^2+10E_0E_e-16E_e^2+3m_e^2}{12m_n^2}$ \\
  $\mathcal{C}_{12}^{\bar{N}}$ &$\frac{(g_{eV}^2+g_{eA}^2)m_e}{E_e}$ &0& $\frac{2(E_0-E_e)}{3(m_n+m_p)}$  & $\frac{2E_0^2+10E_0E_e-15E_e^2+3m_e^2}{6m_n(m_n+m_p)}$   \\
  $\mathcal{C}_{13}^{\bar{N}}$ & $\frac{(g_{eV}^2+g_{eA}^2)m_e}{E_e}$ & 0 &0 & $\frac{2E_0E_e-5E_e^2+3m_e^2}{6m_n(m_n+m_p)}$  \\
  $\mathcal{C}_{14}^{\bar{N}}$ & $\frac{g_{eV}g_{eA}m_e}{E_e}$ &2 & $\frac{2(E_0+5E_e)}{3m_n}$ &   $\frac{2(4E_0E_e+8E_e^2-3m_e^2)}{3m_n^2}$\\
  $\mathcal{C}_{15}^{\bar{N}}$ & $\frac{g_{eV}g_{eA}m_e}{E_e}$& 0  & $-\frac{2(2E_0+E_e)}{3(m_n+m_p)}$ &  $-\frac{2E_0^2+9E_0E_e+E_e^2}{3m_n(m_n+m_p)}$ \\
  $\mathcal{C}_{16}^{\bar{N}}$&$\frac{g_{eV}g_{eA}m_e}{E_e}$ & 0 & 0 & $\frac{-E_eE_0+E_e^2}{3m_n(m_n+m_p)}$  \\
  $\mathcal{C}_{18}^{\bar{N}}$ &$\frac{g_{eV}g_{eA}m_e}{E_e}$ & 0 & 0 & $\frac{2(2E_0E_e-2E_e^2+m_e^2)}{m_n^2}$   \\
  $\mathcal{C}_{22}^{\bar{N}}$ &$\frac{g_{eV}g_{eA}m_e}{E_e}$& 0 & 0 & $\frac{(E_0-E_e)^2}{6m_n(m_n+m_p)}$  \\
  $\mathcal{C}_{23}^{\bar{N}}$ &$\frac{(g_{eV}^2+g_{eA}^2)m_e}{E_e}$ & 0 & 0 & $\frac{2E_0E_e-5E_e^2+3m_e^2}{6m_n(m_n+m_p)}$  \\
  $\mathcal{C}_{24}^{\bar{N}}$&$\frac{g_{eV}g_{eA}m_e}{E_e}$ & 0 & $\frac{8(E_0-E_e)}{3(m_n+m_p)}$ &  $-\frac{2E_0^2-35E_0E_e+39E_e^2-6m_e^2}{3m_n(m_n+m_p)}$ \\
  $\mathcal{C}_{25}^{\bar{N}}$ &$\frac{g_{eV}g_{eA}m_e}{E_e}$& 0 & 0 & $\frac{-2(2E_0^2-5E_e^2+3m_e^2)}{3(m_n+m_p)^2}$  \\
  $\mathcal{C}_{34}^{\bar{N}}$ &$\frac{g_{eV}g_{eA}m_e}{E_e}$& 0 & $\frac{2E_e}{m_n+m_p}$  & $\frac{-5E_0E_e+23E_e^2-6m_e^2}{3m_n(m_n+m_p)}$  \\
  $\mathcal{C}_{35}^{\bar{N}}$ &$\frac{g_{eV}g_{eA}m_e}{E_e}$& 0 & 0 & $-\frac{2(2E_eE_0+E_e^2)}{3(m_n+m_p)^2}$  \\
  $\mathcal{C}_{44}^{\bar{N}}$ &$\frac{(g_{eV}^2+g_{eA}^2)m_e}{E_e}$  & 1 & $\frac{-2(E_0-4E_e)}{3m_n}$ & $-\frac{3E_0^2+26E_0E_e-80E_e^2+15m_e^2}{12m_n^2}$  \\
  $\mathcal{C}_{45}^{\bar{N}}$ &$\frac{(g_{eV}^2+g_{eA}^2)m_e}{E_e}$  & 0 & $-\frac{2(2E_0+E_e)}{3(m_n+m_p)}$ & $\frac{2E_0^2-18E_0E_e-17E_e^2+9m_e^2}{6m_n(m_n+m_p)}$  \\
  $\mathcal{C}_{46}^{\bar{N}}$ &$\frac{(g_{eV}^2+g_{eA}^2)m_e}{E_e}$  & 0 & 0 &  $\frac{-2E_0E_e+5E_e^2-3m_e^2}{6m_n(m_n+m_p)}$ \\
  $\mathcal{C}_{47}^{\bar{N}}$ &$\frac{g_{eV}g_{eA}m_e}{E_e}$& 0 & 0 & $\frac{2(2E_0E_e-2E_e^2+m_e^2)}{m_n^2}$  \\
  $\mathcal{C}_{48}^{\bar{N}}$ &$\frac{(g_{eV}^2+g_{eA}^2)m_e}{E_e}$ &0 &0 & $\frac{2(2E_0E_e-2E_e^2+m_e^2)}{m_n^2}$\\
  $\mathcal{C}_{55}^{\bar{N}}$ &$\frac{(g_{eV}^2+g_{eA}^2)m_e}{E_e}$  &0& 0 &  $\frac{E_0^2+2E_0E_e}{3(m_n+m_p)^2}$  \\
\hline\hline
\end{tabular}
\caption{Analytical results for $\mathcal{C}_{ij}^{\bar{N}}$, where $L_{ij}^{\bar{N}}$ refers to the factor from the lepton part. Some contributions such as $\mathcal{C}_{17}^{\bar{N}}$ and $\mathcal{C}_{26}^{\bar{N}}$ are zero and have been omitted in the table.}
\label{table:Nbar-ana}
\end{table}
\begin{table}[htbp]
\centering
\begin{tabular}{p{1cm}p{2cm} p{2cm} p{3cm} p{6cm}}
  \hline\hline
      &  $L_{ij}^{\bar{Q}}$   & $\mathcal{O}(1)$ & $\mathcal{O}(m_n^{-1})$ &$\mathcal{O}(m_n^{-2})$ \\
\hline $\mathcal{C}_{11}^{\bar{Q}}$ &  $g_{eV}^2+g_{eA}^2$ &    0   &    $\frac{E_0-E_e}{3m_n}$   & $\frac{(3E_0^2+8E_0E_e-8E_e^2)+(-2E_0+8E_e+3m_e)m_e}{12m_n^2}$    \\
$\mathcal{C}_{12}^{\bar{Q}}$ & $g_{eV}^2+g_{eA}^2$  & 0  &    $\frac{2(E_0-E_e)}{3(m_n+m_p)}$   &   $\frac{2(E_0^2+2E_0E_e)+(-6E_0+15E_e+3m_e)m_e}{6m_n(m_n+m_p)}$  \\
$\mathcal{C}_{13}^{\bar{Q}}$ & $g_{eV}^2+g_{eA}^2$ &   0    &    0   &  $-\frac{(2E_0-5E_e-3m_e)m_e}{6m_n(m_n+m_p)}$   \\
$\mathcal{C}_{14}^{\bar{Q}}$ & $g_{eV}g_{eA}$ &  $2$   &    $\frac{2(E_0+2E_e-3m_e)}{3m_n}$   & $\frac{16(E_0E_e-E_e^2)+2(4E_0-16E_e-3m_e)m_e}{3m_n^2}$    \\
$\mathcal{C}_{15}^{\bar{Q}}$ &   $g_{eV}g_{eA}$ &  0   &   $\frac{2(-2E_0+2E_e+3m_e)}{3(m_n+m_p)}$   & $-\frac{(2E_0^2+8E_0E_e-16E_e^2)+(E_0+17E_e)m_e}{3m_n(m_n+m_p)}$    \\
$\mathcal{C}_{16}^{\bar{Q}}$ &  $g_{eV}g_{eA}$    &   0   &0 &   $\frac{(E_0-E_e)m_e}{3m_n(m_n+m_p)}$  \\
$\mathcal{C}_{18}^{\bar{Q}}$ &$g_{eV}g_{eA}$&    0   &   0    &  $\frac{8(E_0E_e-E_e^2)+2(-2E_0+2E_e+3m_e)m_e}{3m_n^2}$   \\
$\mathcal{C}_{22}^{\bar{Q}}$ & $g_{eV}^2+g_{eA}^2$ &   0    &   0     &  $\frac{(E_0^2-6E_0E_e+8E_e^2)-(4E_0-7E_e)m_e}{6m_n(m_n+m_p)}$   \\
$\mathcal{C}_{23}^{\bar{Q}}$ & $g_{eV}^2+g_{eA}^2$ & 0    &    0   &  $-\frac{(2E_0-5E_e-3m_e)m_e}{6m_n(m_n+m_p)}$   \\
$\mathcal{C}_{24}^{\bar{Q}}$& $g_{eV}g_{eA}$ &   0    &  $\frac{4(2E_0-5E_e-3m_e)}{3(m_n+m_p)}$     & $\frac{-2(E_0^2-28E_0E_e+48E_e^2)+(21E_0-57E_e+6m_e)m_e}{3m_n(m_n+m_p)}$    \\
$\mathcal{C}_{25}^{\bar{Q}}$ &$g_{eV}g_{eA}$ &   0    &  0     &  $\frac{-4(E_0^2-2E_0E_e-2E_e^2)+2(4E_0-E_e-3m_e)m_e}{3(m_n+m_p)^2}$   \\
$\mathcal{C}_{34}^{\bar{Q}}$ &$g_{eV}g_{eA}$ &  0   &  $\frac{-2m_e}{m_n+m_p}$     & $\frac{(5E_0-23E_e-6m_e)m_e}{3m_n(m_n+m_p)}$    \\
$\mathcal{C}_{35}^{\bar{Q}}$ &$g_{eV}g_{eA}$ &   0   &  0     &  $\frac{2(2E_0+E_e)m_e}{3(m_n+m_p)^2}$   \\
$\mathcal{C}_{44}^{\bar{Q}}$ &   $g_{eV}^2+g_{eA}^2$  & 1    &   $\frac{-2E_0+11E_e+3m_e}{3m_n}$    &$-\frac{(3E_0^2+40E_0E_e-136E_e^2)+(14E_0-56E_e+15m_e)m_e}{12m_n^2}$     \\
$\mathcal{C}_{45}^{\bar{Q}}$& $g_{eV}^2+g_{eA}^2$&   0    &  $-\frac{2(2E_0+E_e)}{3(m_n+m_p)}$     & $\frac{2(E_0^2-14E_0E_e-8E_e^2)+(-10E_0+E_e+9m_e)m_e}{6m_n(m_n+m_p)}$    \\
$\mathcal{C}_{46}^{\bar{Q}}$& $g_{eV}^2+g_{eA}^2$&  0 & 0 & $\frac{(2E_0-5E_e-3m_e)m_e}{6m_n(m_n+m_p)}$ \\
$\mathcal{C}_{47}^{\bar{Q}}$ &$g_{eV}g_{eA}$ &   0    &  0     & $\frac{8(E_0E_e-E_e^2)+2(-2E_0+2E_e+3m_e)m_e}{3m_n^2}$    \\
$\mathcal{C}_{48}^{\bar{Q}}$& $g_{eV}^2+g_{eA}^2$&   0    &   0    &  $\frac{16(E_0E_e-E_e^2)+2(2E_0-2E_e+3m_e)m_e}{3m_n^2}$   \\
$\mathcal{C}_{55}^{\bar{Q}}$ &$g_{eV}^2+g_{eA}^2$ &   0    &   0    & $\frac{E_0^2+2E_0E_e}{3(m_n+m_p)^2}$    \\
\hline\hline
\end{tabular}
\caption{Analytical results for $\mathcal{C}_{ij}^{\bar{Q}}$, where $L_{ij}^{\bar{Q}}$ refers to the factor from the lepton part. Some contributions such as $\mathcal{C}_{17}^{\bar{Q}}$ and $\mathcal{C}_{26}^{\bar{Q}}$ are zero and have been omitted in the table.}
\label{table:Qbar-ana}
\end{table}

The coupling constants $f_1,f_4$ are just the weak coupling constants $g_V$ and $g_A$ with $f_1\equiv g_V$ and $f_4\equiv g_A$, and the ratio $f_4/f_1\equiv \lambda$ can be determined from \cite{lambda-Ex}. If assuming the $SU(2)$ symmetry and using CVC hypothesis, one has $f_1=1$ and $f_2\equiv\kappa=\kappa_p-\kappa_n$ where $\kappa_p$ and $\kappa_n$ are the anomalous magnetic moments of proton and the neutron \cite{PDG-2020}. The scalar and weak electricity coupling constants $f_3=0$ and $f_5=0$ corresponding to no SCC \cite{SCC}. The pseudoscalar coupling $f_6$ is induced by the strong interaction effects, with the partially conserved axial vector current (PCAC) hypothesis $f_6=\frac{4m_n^2f_4}{m_{\pi}^2}$  \cite{FFs-B}.

Combined the conditions $f_1=1,f_2=\kappa,f_4=\lambda, f_3=f_5=f_6=f_7=f_8=0$ and $ g_{eV}=g_{eA}=1$ with the expressions of $\mathcal{O}(1)$ in Tables \ref{table:zeta-ana}-\ref{table:Qbar-ana},
one can get the ``bare'' correlation coefficients given by the Eq. (1) of Ref. \cite{Ivanov-2017}. We want to mention that a global factor $1+3\lambda^2$ has been pulled into the ``bare'' results to guarantee the same definition of $\zeta(E_e)$. {\color{black} Furthermore, considering our expressions of $\mathcal{O}(m_n^{-1})$, as well as the replacement $\frac{1}{m_n}\rightarrow \frac{2}{m_n+m_p}$, one can completely reproduce the results $\zeta(E_e), \zeta(E_e)A_w(E_e), \zeta(E_e)G(E_e) $ in Eqs. (6,7) of Ref. \cite{Ivanov-2017}. Meanwhile, the differences for the other two results are expressed as  $\bar{N}(E_e)-\zeta(E_e)N(E_e)=\frac{2(E_0-2E_e)m_e\lambda}{E_e(m_n+m_p)}$ and $ \bar{Q}(E_e)-\zeta(E_e)Q(E_e)=\frac{2(E_0-2E_e-m_e)\lambda}{m_n+m_p}$.} Actually, although the results of $\mathcal{O}(m_n^{-1})$ can be reproduced partly, we wanted to emphasize that the phase factor $\beta_{\RM1}$ is more reasonable than $\beta_{\RM2}$.

\section{Numerical Comparison and Discussions}
In the literatures, the expansion on $m_n^{-1}$(or $m_N^{-1}$) usually is done for the phase space factor and the amplitude independently, in this work we also carry the above calculation with $\beta_{\RM1}$ and $\beta_{\RM2}$ as inputs, and we do not show the corresponding analytic expressions $X_{\RM1,\RM2}(E_e)$ here, but just show the numerical comparison.

In the numerical comparison, we take $m_n=939.56542 \text{ MeV}$, $m_p=938.27209 \text{ MeV}$, $m_e=0.51100 \text{ MeV}$\cite{PDG-2020}, and the neutrino mass $m_v\approx 0$.

\begin{figure*}[htb]
\centering
\includegraphics[height=6cm]{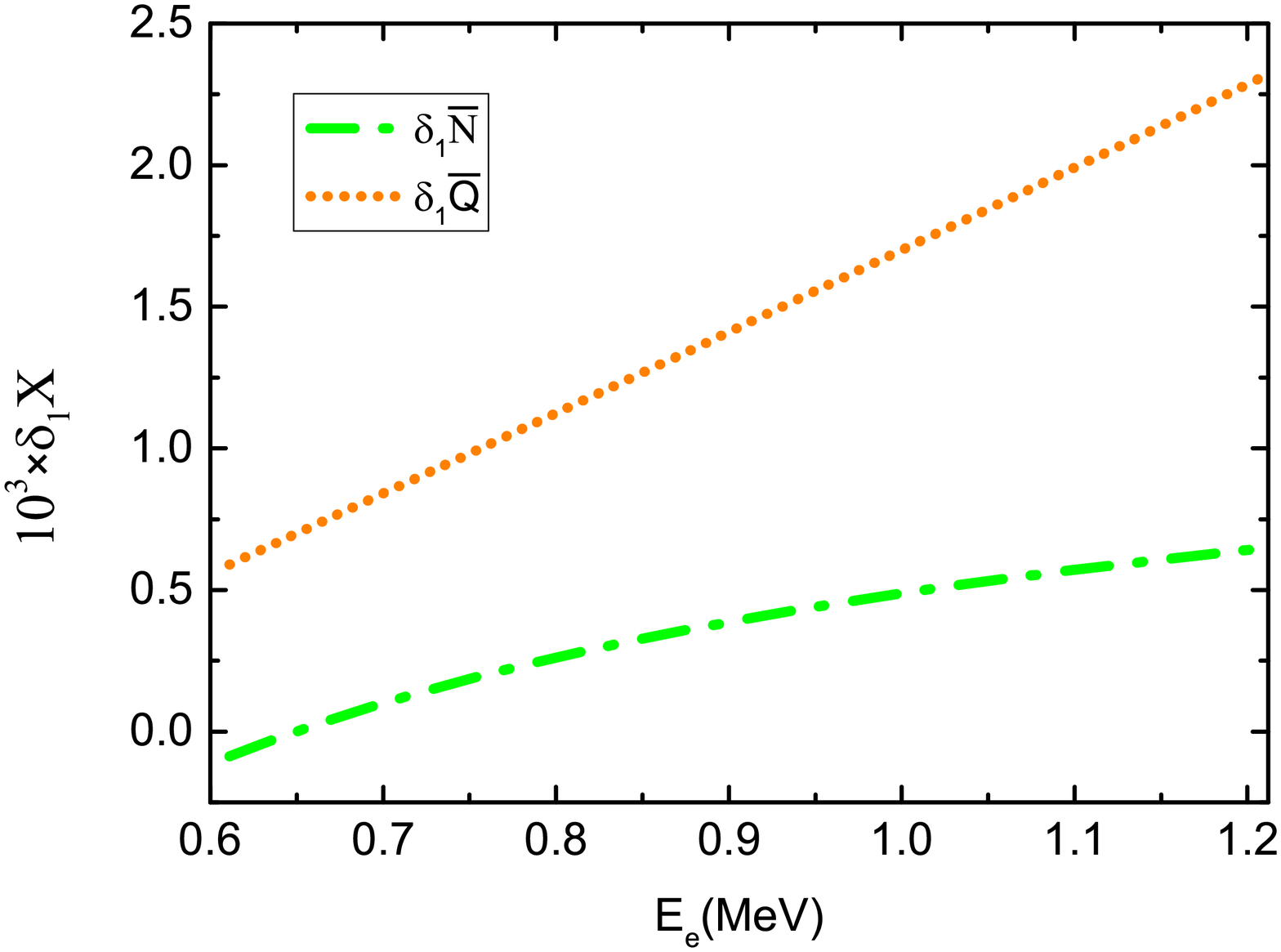}\includegraphics[height=6cm]{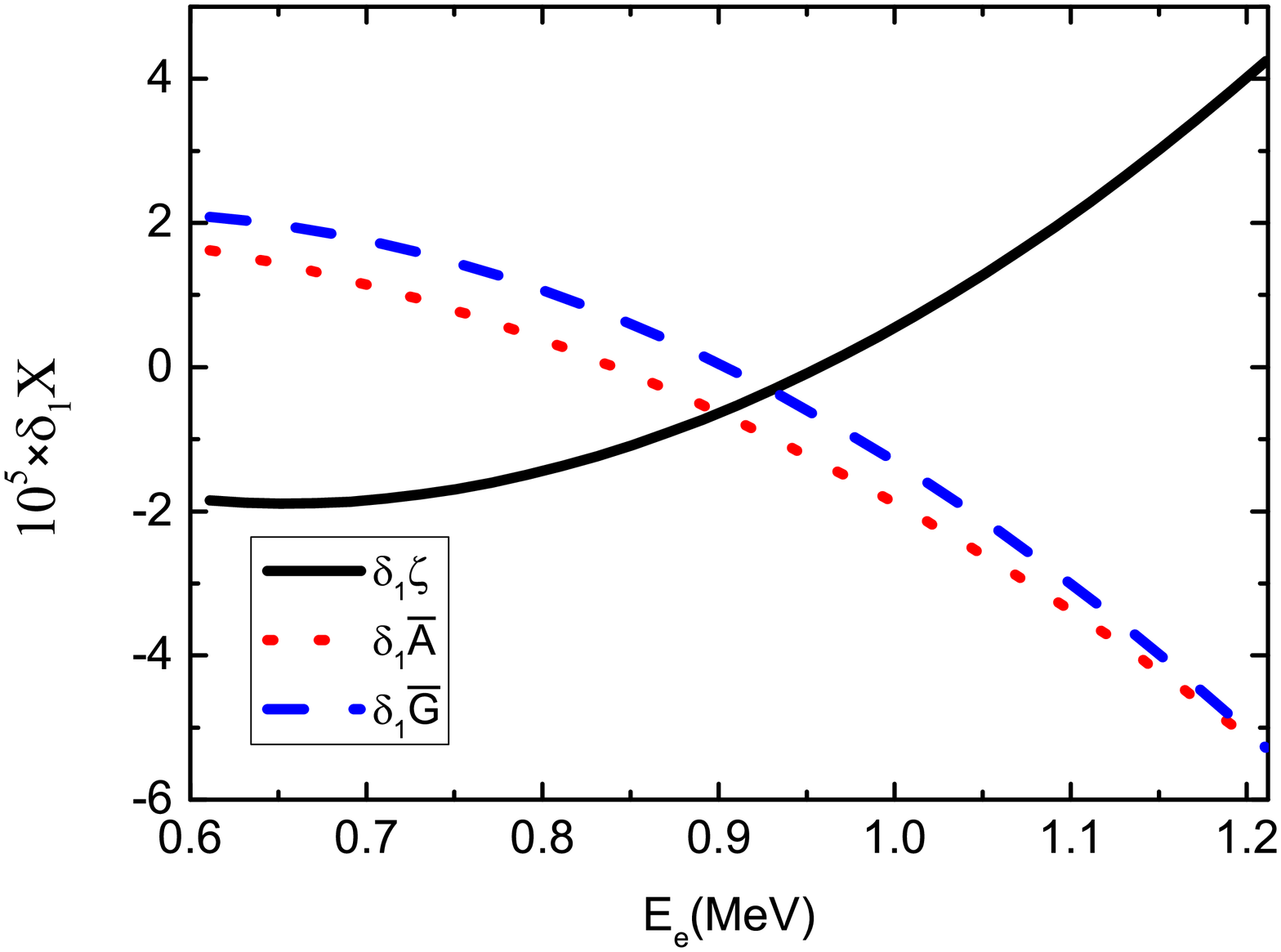}
\caption{Numeric results for $\delta_1X$ {\it vs.} $E_e$, where the left panel is the result for $X=\bar{N},\bar{Q}$ and the right panel is the result for $X=\zeta, \bar{A},\bar{G}$. }
\label{figure:delta-1}
\end{figure*}

In Ref. \cite{Ivanov-2017}, the phase factor is taken as $\beta_{\RM2}$ and the expansion of ${\cal M}$ is applied to obtain the relevant correlation coefficients. The authors finally get the analytic expressions shown as the Eqs. (6, 7) in Ref. \cite{Ivanov-2017}. To compare the results in a direct way, firstly we also take $\beta_{\RM2}$ as input and expand the full result $d^5\sigma_n(E_e,\vec{k}_e,\vec{\xi}_e, \vec{\xi}_n)$ to get the expressions. We define the difference of these two results as
\begin{eqnarray}
\delta_{1} X&\equiv& X^{\text{our}}_{\beta_{\RM2},A} - X^{\text{ref}}_{\beta_{\RM2},A},
\end{eqnarray}
where $X^{\text{ref}}_{\beta_{\RM2},A}$ refers to the results by the Eqs. (6, 7) in Ref. \cite{Ivanov-2017}, $X^{\text{our}}_{\beta_{\RM2},A}$ refers to the results by our calculation with $\beta_{\RM2}$ as input, the subindex $A$ refers to the choice of the parameters as Ref. \cite{Ivanov-2017}:
\begin{eqnarray}
f_1=1, f_2=3.7058, f_4=-1.2767, f_3=f_5=f_6=f_7=f_8=0, g_{eV}=g_{eA}=1,
\label{Eq:parametersA}
\end{eqnarray}
The $E_e$ dependences of $\delta_{1} X$ are presented in Fig. \ref{figure:delta-1} where the (orange) short dotted curve and the (green) dash-dotted curve in the left panel refer to the results of $\delta_{1} \bar{Q}$ and $\delta_{1} \bar{N}$, the (black) solid curve, the (red) dotted curve and the (blue) dashed curve in the right panel are associated with $\delta_{1} \zeta,\delta_{1} \bar{A},\delta_{1} \bar{G}$, respectively. The results clearly show that the differences $\delta_{1} \zeta$, $\delta_{1} \bar{A}$ and $\delta_{1} \bar{G}$ are at the order $10^{-5}$,  the difference $\delta_{1} \bar{N}$ is at the order $10^{-4}$ and the difference $\delta_{1} \bar{Q}$ even reaches the order $10^{-3}$ at large $E_e$. The numerical comparison shows the differences are not negligible.

\begin{figure}[htbp]
\centering
\includegraphics[height=8cm]{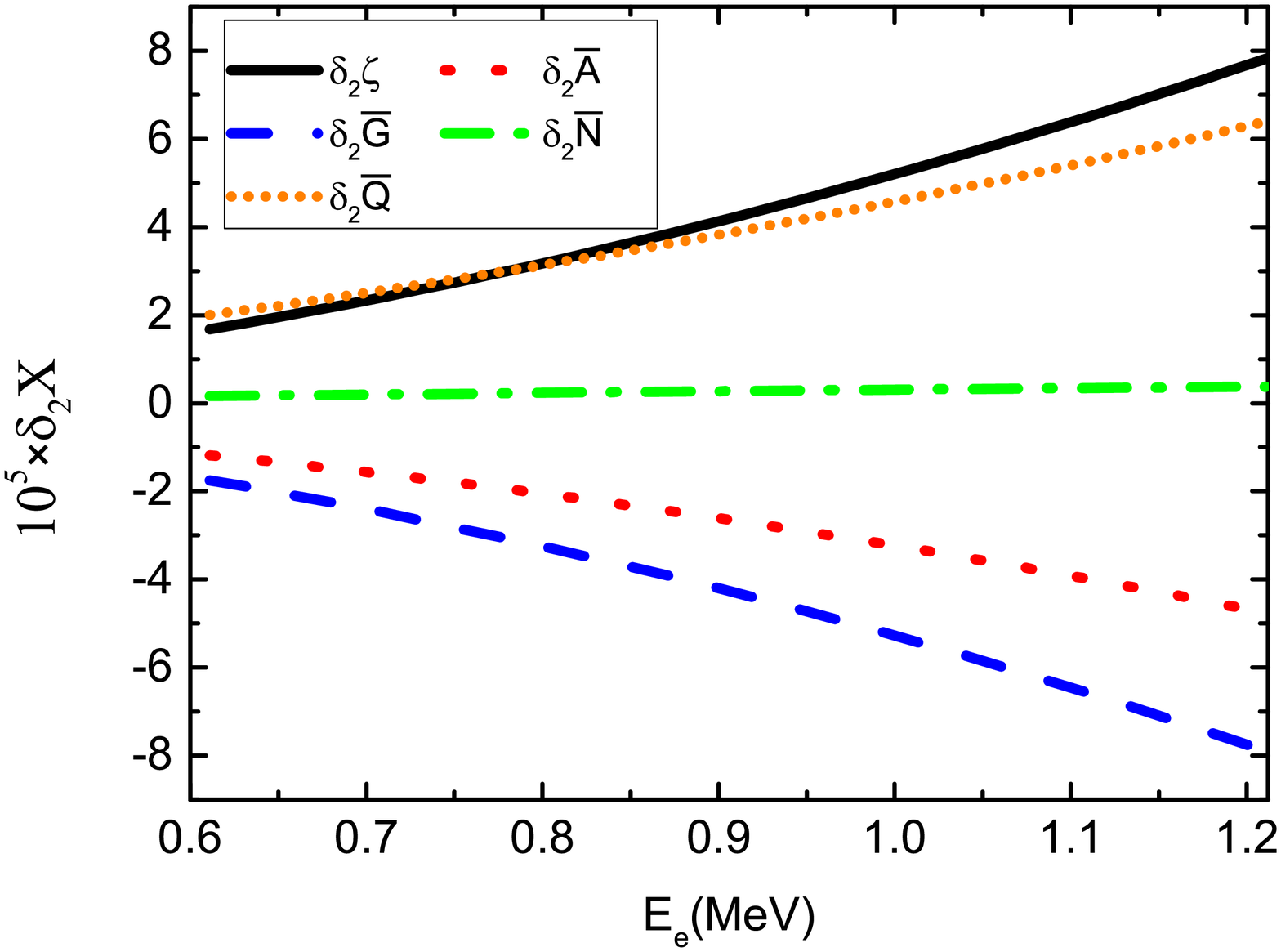}
\caption{Numeric results for $\delta_2X$ {\it vs.} $E_e$ where the index $X$ refers to $\zeta, \bar{A},\bar{G},\bar{N},\bar{Q}$, respectively.}
\label{figure:delta-2}
\end{figure}

In our calculation, the analytic expressions are got by taking the phase space factor $\beta$ as input. To show the difference due to the different choices of the phase factor, we define
\begin{eqnarray}
\delta_{2} X&\equiv& X_{\beta,\text{A}}^{\text{our}} - X_{\beta_{\RM1},\text{A}}^{\text{our}},
\end{eqnarray}
where the index ``our'' refers to our calculation, ``$\beta, \beta_{\RM1}$'' refer to the input phase factors, and the index ``A" refers to the choice of the paramters $f_i$ and $g_{eV, eA}$ as the Eq. (\ref{Eq:parametersA}). The $E_e$ dependence of $\delta_2 X$ is presented in Fig. \ref{figure:delta-2} where the definitions of the curves are the same as those in Fig. \ref{figure:delta-1}. The results show an interesting property that the absolute magnitudes of $\delta_2\zeta, \delta_2\bar{A},\delta_2\bar{G},\delta_2\bar{Q}$ increase when $E_e$ increases and are at order of $10^{-5}$, while $\delta_2\bar{N}$ are not sensitive to $E_e$ and at the order of $10^{-6}$.

Furthermore, to consider the contributions at the order $10^{-5}$, the contributions from the parameters $f_6,f_7,f_8$ may also play their roles. To show these contributions, we define the difference
\begin{eqnarray}
\delta_3 X&=& X_{\beta,\text{B}}^{\text{our}} - X_{\beta,\text{A}}^{\text{our}},
\end{eqnarray}
where the subindex ``B'' refers to the case with nonzero $f_{6,7,8}$ and same $f_{1,2,3,4,5}$ with case ``A". The values of $f_{6,7,8}$ are taken from Ref. \cite{FFs-B}  as
\begin{eqnarray}
f_6=228, ~f_7=2.5f_1,~f_8=1.92f_4.
\label{Eq:parametersB}
\end{eqnarray}
Here we want to mention that our $f_6$ are twice of the $f_6$ in Ref. \cite{FFs-B} due to the different definitions of the form factors in $\Gamma_{Wnp}$.

\begin{figure}[htbp]
\centering
\includegraphics[height=8cm]{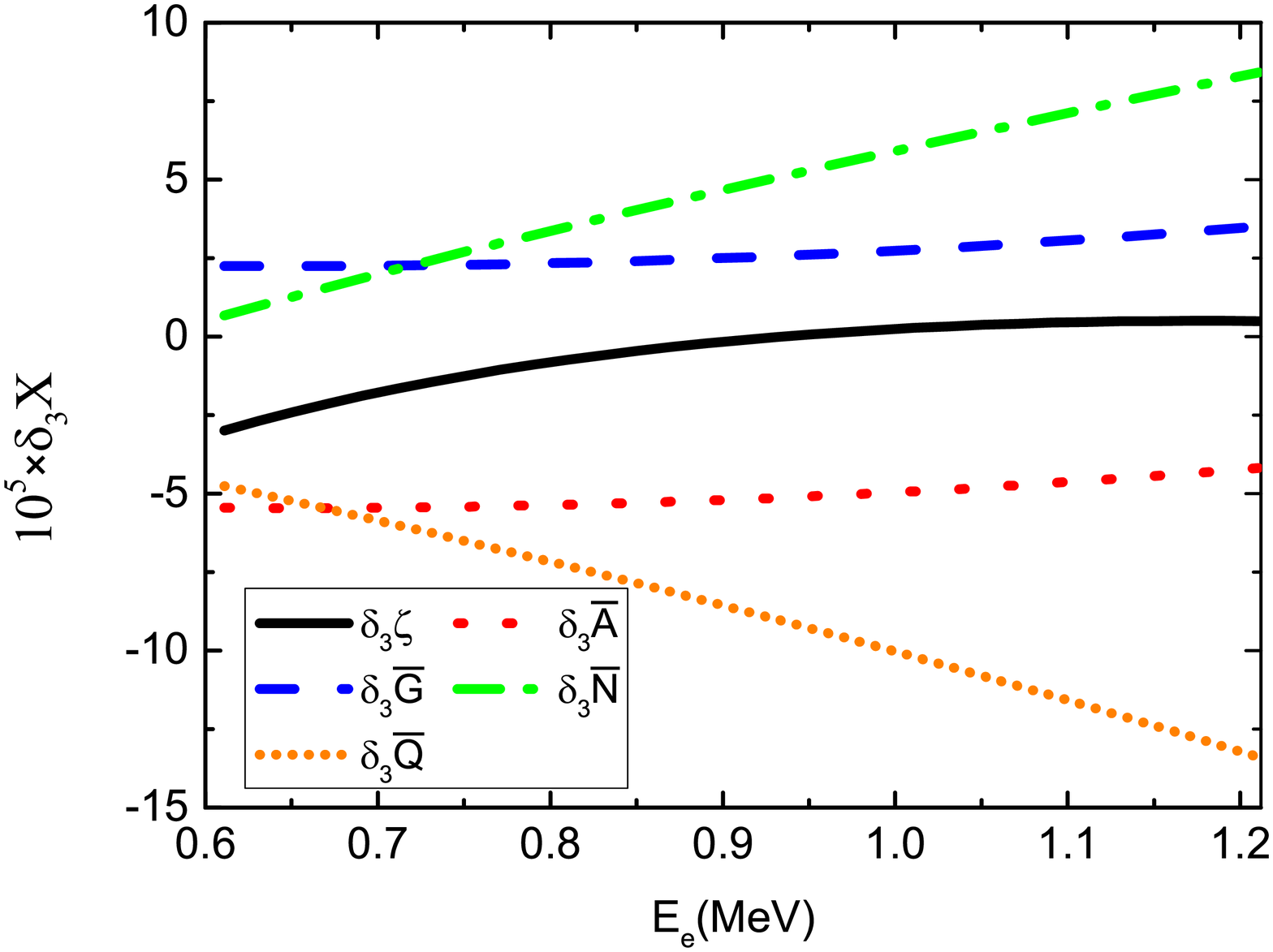}
\caption{Numeric results for $\delta_3X$ {\it vs.} $E_e$ where the index $X$ refers to $\zeta, \bar{A},\bar{G},\bar{N},\bar{Q}$, respectively.}
\label{figure:delta-3}
\end{figure}

In Fig. \ref{figure:delta-3}, we present the $E_e$ dependence of $\delta_3 X$ where the definitions of the curves are the same with those in Fig. \ref{figure:delta-1}. One can find that the absolute magnitudes of $\delta_3\zeta$, $\delta_3\bar{A}$, $\delta_3\bar{G}$ and $\delta_3 \bar{N}$ are at the order $10^{-5}$, and the magnitude of $\delta_3\bar{Q}$ even reaches $10^{-4}$ at large $E_e$. These properties shows that the contributions from $f_6, f_7$ and $f_8$ should also be considered when aiming to the precision $10^{-5}$.

Combing the above detailed numerical comparisons, we can see that our results are very different from the results Eqs. (6,7)  in Ref. \cite{Ivanov-2017} even the same phase space factor $\beta_{\RM2}$ is taken as input, especially for the $\bar{Q}$ case. Our results clearly show that the full phase space factor and the other form factors should be kept in the estimation to reach the $10^{-5}$ precision. Also the expansion on $m_n^{-1}$ should be done carefully. In the practical calculation, we also check our results to higher order such as  $O(m_n^{-4})$ and find the results are almost the same.

\section{Acknowledgments}
This work is supported by the  National Natural Science Foundations of China under Grant No. 12075058, No.12047503 and No. 11975075. Hui-Yun Cao was supported by the Science and Technology Research Project of Hubei Provincial Education Department (Grants no. Q20222502).

\section{Appendix}

\subsection{Kinematics}
In this Appendix, we list the manifest expressions for the
momenta used in the calculation. In the rest frame of neutron, the momenta and the corresponding spin vectors of the polarized neutron and the polarized electron are taken as
\begin{eqnarray}
p_n&\equiv& (m_n,\vec{0}), \nonumber\\
p_e&\equiv& (E_e,\vec{k}_e),\nonumber\\
p_v &\equiv& (E_v,\vec{k}_v),\nonumber\\
p_p &=&p_n-p_e-p_v=(m_n-E_e-E_v,-\vec{k}_e-\vec{k}_v),\nonumber\\
S_n&=&(0,\vec{\xi}_n), \nonumber\\ S_e&=&(\frac{\vec{k}_e\cdot\vec{\xi}_e}{m_e},\vec{\xi}_e+\frac{\vec{k}_e(\vec{k}_e\cdot\vec{\xi}_e)}{m_e(E_e+m_e)}),
\end{eqnarray}

Furthermore, after using the on-shell condition $p_p^2=m_p^2$, one can obtain
\begin{eqnarray}
E_v &=&\frac{m_n(E_0-E_e)}{m_n-E_e+\vec{k}_e \cdot\vec{n}_v},
\end{eqnarray}
where the unit vectors $\vec{n}_v$ is directed along the neutrino three-momentum $\vec{k}_v$. This is just the first replacement rule in Eq. (\ref{Eq: replace-rules}).

\subsection{The integration over the solid angle of neutrino}
After the expansion on $m_n^{-1}$, the integration of $d\Omega_v$ can be done using the following results:
\begin{eqnarray}
\int (\vec{n}_v\cdot \vec{a}_1) d\Omega_v &=& 0, \nonumber\\
\int (\vec{n}_v\cdot \vec{a}_1) (\vec{n}_v\cdot \vec{a}_2) d\Omega_v &=& \frac{4\pi}{3}\vec{a}_1\cdot\vec{a}_2, \nonumber\\
\int (\vec{n}_v\cdot \vec{a}_1) (\vec{n}_v\cdot \vec{a}_2)(\vec{n}_v\cdot \vec{a}_3) d\Omega_v &=& 0, \nonumber\\
\int (\vec{n}_v\cdot \vec{a}_1) (\vec{n}_v\cdot \vec{a}_2)(\vec{n}_v\cdot \vec{a}_3)(\vec{n}_v\cdot \vec{a}_4)d\Omega_v &=& \frac{4\pi}{15}\Big[ (\vec{a}_1\cdot\vec{a}_2)(\vec{a}_3\cdot\vec{a}_4)+ (\vec{a}_1\cdot\vec{a}_3) (\vec{a}_2\cdot\vec{a}_4)\nonumber\\
&&+ (\vec{a}_1\cdot\vec{a}_4)(\vec{a}_2\cdot\vec{a}_3)\Big],  \nonumber\\
\int (\vec{n}_v\cdot \vec{a}_1) (\vec{n}_v\cdot \vec{a}_2)(\vec{n}_v\cdot \vec{a}_3)(\vec{n}_v\cdot \vec{a}_4)(\vec{n}_v\cdot \vec{a}_5)d\Omega_v &=& 0,
\end{eqnarray}
and
\begin{eqnarray}
&&\int (\vec{n}_v\cdot \vec{a}_1) (\vec{n}_v\cdot \vec{a}_2)(\vec{n}_v\cdot \vec{a}_3)(\vec{n}_v\cdot \vec{a}_4)(\vec{n}_v\cdot \vec{a}_5)(\vec{n}_v\cdot \vec{a}_6) d\Omega_v \nonumber\\ &=&\frac{4\pi}{105}\Big[(\vec{a}_1\cdot\vec{a}_2)(\vec{a}_3\cdot\vec{a}_4)(\vec{a}_5\cdot\vec{a}_6) +(\vec{a}_1\cdot\vec{a}_3)(\vec{a}_2\cdot\vec{a}_4)(\vec{a}_5\cdot\vec{a}_6) + (\vec{a}_1\cdot\vec{a}_4)(\vec{a}_2\cdot\vec{a}_3)(\vec{a}_5\cdot\vec{a}_6)  \nonumber\\
&& + (\vec{a}_1\cdot\vec{a}_2)(\vec{a}_3\cdot\vec{a}_5)(\vec{a}_4\cdot\vec{a}_6) + (\vec{a}_1\cdot\vec{a}_3)(\vec{a}_2\cdot\vec{a}_5)(\vec{a}_4\cdot\vec{a}_6) + (\vec{a}_1\cdot\vec{a}_5)(\vec{a}_2\cdot\vec{a}_3)(\vec{a}_4\cdot\vec{a}_6) \nonumber\\
&& + (\vec{a}_1\cdot\vec{a}_2)(\vec{a}_3\cdot\vec{a}_6)(\vec{a}_4\cdot\vec{a}_5) + (\vec{a}_1\cdot\vec{a}_3)(\vec{a}_2\cdot\vec{a}_6)(\vec{a}_4\cdot\vec{a}_5) + (\vec{a}_1\cdot\vec{a}_6)(\vec{a}_2\cdot\vec{a}_3)(\vec{a}_4\cdot\vec{a}_5) \nonumber\\
&& + (\vec{a}_1\cdot\vec{a}_4)(\vec{a}_2\cdot\vec{a}_5)(\vec{a}_3\cdot\vec{a}_6) + (\vec{a}_1\cdot\vec{a}_5)(\vec{a}_2\cdot\vec{a}_4)(\vec{a}_3\cdot\vec{a}_6) + (\vec{a}_1\cdot\vec{a}_4)(\vec{a}_2\cdot\vec{a}_6)(\vec{a}_3\cdot\vec{a}_5) \nonumber\\
&& + (\vec{a}_1\cdot\vec{a}_6)(\vec{a}_2\cdot\vec{a}_4)(\vec{a}_3\cdot\vec{a}_5)+ (\vec{a}_1\cdot\vec{a}_5)(\vec{a}_2\cdot\vec{a}_6)(\vec{a}_3\cdot\vec{a}_4)+ (\vec{a}_1\cdot\vec{a}_6)(\vec{a}_2\cdot\vec{a}_5)(\vec{a}_3\cdot\vec{a}_4)  \Big]. \nonumber\\
\end{eqnarray}
where $\vec{a}_i$ are some vectors independent on $\Omega_v$.

\end{document}